\documentclass[prc,nofootinbib]{revtex4-2}

\pdfoutput=1

\usepackage[utf8]{inputenc}
\usepackage{amsmath,amssymb,slashed}

\usepackage{bbm}
\usepackage{graphicx,xcolor}

\usepackage{hyperref}

\newcommand{\ie}{{\it i.e.}}

\begin{document}

\title{Chiral perturbation theory}

\author{S. Scherer}
\affiliation{ Institut f\"ur Kernphysik, Johannes Gutenberg-Universit\"at Mainz, D-55099 Mainz, Germany}

\author{M. R. Schindler}
\affiliation{Department of Physics and Astronomy, University of South Carolina, Columbia, SC, 29208, USA}

\begin{abstract}
Chiral perturbation theory (ChPT) is an effective field theory that describes the properties of strongly-interacting systems at energies far below typical hadron masses. 
The degrees of freedom are hadrons instead of the underlying quarks and gluons. 
ChPT is a systematic and model-independent appro\-xi\-ma\-tion method based on an expansion of amplitudes in terms of light-quark masses and momenta.
The following is a brief overview of ChPT that is largely based on Ref.~\cite{Scherer:2012xha}, which can be referred to for a more detailed introduction.
\end{abstract}

\maketitle

\section{QCD and chiral symmetry}
    The QCD Lagrangian---obtained by applying the gauge principle with
respect to the $\text{SU}(3)$ color group to the free Lagrangians of six quark flavors with masses $m_f$---reads
\begin{equation}
    \label{chpt:lqcd}
    {\cal L}_{\rm QCD}=\sum_{f=u,\ldots,t}
    \bar{q}_f\left(i\slashed{\bf D}-m_f\right)q_f
    -\frac{1}{2}\text{Tr}_c\left({\bf F}_{\mu\nu}{\bf F}^{\mu\nu}\right).
\end{equation}
    For each quark flavor $f$, the quark field $q_f$ is a color triplet,
transforming in the triplet representation,
\begin{equation}
    \label{chpt:qtrans}
    q_f(x)\mapsto U(x)q_f(x),   
\end{equation}
where $U(x)$ denotes a smooth space-time-dependent SU(3) matrix.
Using the Gell-Mann matrices \cite{Gell-Mann:1961omu}, the eight gluon fields ${\cal A}^A_{\mu}$ are collected in a traceless,
Hermitian, $3\times 3$ matrix ${\bf A}_\mu=\lambda^A {\cal A}^A_{\mu}/2$ (summation over repeated indices implied), transforming inhomo\-geneously under a gauge transformation,
\begin{equation}
    \label{chpt:atrans}
    {\bf A}_\mu(x)\mapsto U (x){\bf A}_\mu(x)U^\dagger(x)
    +\frac{i}{g_s}\partial_\mu U(x) U^\dagger(x),
\end{equation}
where $g_s$ denotes the SU(3) gauge coupling constant.
    In terms of ${\bf A}_\mu$, the covariant derivative of the quark fields is 
defined as
\begin{equation}
    \label{chpt:dmuqf}
    {\bf D}_\mu q_f=\left(\partial_\mu+ig_s{\bf A}_\mu\right)q_f.
\end{equation}
    Finally, the field strength tensor is given by
\begin{equation}
    {\bf F}_{\mu\nu}=\partial_\mu{\bf
A}_\nu-\partial_\nu{\bf A}_\mu +ig_s[{\bf A}_\mu,{\bf A}_\nu].
\end{equation}
    By construction, the Lagrangian of Eq.~\eqref{chpt:lqcd} is invariant under 
the combined transformations of Eqs.~\eqref{chpt:qtrans} and \eqref{chpt:atrans}.
From the point of view of gauge invariance, the strong-interaction Lagrangian could also involve a term of the type \cite{Gasser:1984gg}
\begin{equation}
\label{chpt:ltheta} 
    {\cal L}_\theta=\frac{g^2_s\bar{\theta}}{32\pi^2}
    \epsilon_{\mu\nu\rho\sigma} \text{Tr}_c
    \left({\bf F}^{\mu\nu}{\bf F}^{\rho\sigma}\right),\quad \epsilon_{0123}=1,
\end{equation}
where $\epsilon_{\mu\nu\rho\sigma}$ denotes the totally antisymmetric Levi-Civita tensor.
The so-called $\theta$ term of Eq.~\eqref{chpt:ltheta} implies an explicit $P$ and $CP$ violation of the strong interactions.
The present empirical information on the neutron elec\-tric dipole moment \cite{Abel:2020pzs} indicates that the $\theta$ term is small and, in the following, we will omit Eq.~\eqref{chpt:ltheta} from our discussion.

Since the covariant derivative of the quark fields is flavor independent,
the Lagrangian of Eq.~\eqref{chpt:lqcd} has additional, accidental, and in this case global, symme\-tries aside from the gauge symmetry.
Both the dynamics of the theory (via spontaneous symmetry breaking) and the values of the quark masses impact how these symme\-tries are (approximately) realized in nature.
Dynamical chiral symmetry breaking introduces the scale $\Lambda_\chi=4\pi F_0$ (see below) of the order of 1 GeV \cite{Manohar:1983md}.
In this context it is common to divide the six quark flavors into the three light quarks $u$, $d$, and $s$ with $m_l<\Lambda_\chi$ and the three heavy flavors $c$, $b$, and $t$ with $m_h>\Lambda_\chi$.
    As a theoretical starting point, one may consider two limits, namely,
sending the light-quark masses to zero (chiral limit) and the heavy-quark masses to infinity.
    In Ref.~\cite{Leutwyler:2000cw}, this situation is referred to as a
``theoretician's paradise.''   
    In the following, we exclusively concentrate on the chiral limit for either
two ($u,d$) or three ($u,d,s)$ light quarks and omit the heavy quarks from our dis\-cussion.
    Introducing left-handed and right-handed quark fields (color and flavor 
indices omitted) as
\begin{equation}
q_L=\frac{1}{2}\left({\mathbbm 1}-\gamma_5\right)q,\,\,
q_R=\frac{1}{2}\left({\mathbbm 1}+\gamma_5\right)q,\,\,
\gamma_5=i\gamma^0\gamma^1\gamma^2\gamma^3,
\end{equation}
the QCD Lagrangian in the chiral limit decomposes into
\begin{eqnarray}
    {\cal L}^0_{\rm QCD}&=&\sum_{l=u,d,s}
    \left(\bar{q}_{L,l}i\slashed{\bf D}q_{L,l}+\bar{q}_{R,l}i\slashed{\bf D}
    q_{L,R}\right)
    -\frac{1}{2}\text{Tr}_c\left({\bf F}_{\mu\nu}{\bf F}^{\mu\nu}\right).
    \label{chpt:lqcd0lr} 
\end{eqnarray}
    In the massless limit, the helicity of a quark is a good quantum number 
which is conserved in the interaction with gluons.
    Moreover, the classical Lagrangian in the chiral limit has a global
$\text{U}(3)_L\times\text{U}(3)_R$ symmetry, \ie, it is invariant under independent unitary flavor trans\-form\-ations of the left-handed and the right-handed quark fields,
\begin{displaymath}
    q_L\mapsto U_L q_L,\quad q_R\mapsto U_R q_R.
\end{displaymath}
    At the classical level, this chiral symmetry results in $2\times(8+1)=18$ conserved currents:
\begin{align*}
L^\mu_a&=\bar{q}_L\gamma^\mu\frac{\lambda_a}{2}q_L, &&
R^\mu_a=\bar{q}_R\gamma^\mu\frac{\lambda_a}{2}q_R, && a=1,\ldots,8,\\
V^\mu&=\bar q_R\gamma^\mu q_R+\bar{q}_L\gamma^\mu q_L,
&& A^\mu=\bar q_R\gamma^\mu q_R-\bar{q}_L\gamma^\mu q_L.
\end{align*}
Here, the Gell-Mann matrices act in flavor space, since $q_R$ and $q_L$ are flavor triplets.\footnote{Lower case Roman letters denote SU(3) flavor indices.}
Because of quantum effects the singlet axial-vector current $A^\mu=\bar q \gamma^\mu\gamma_5 q$ develops a so-called anomaly, resulting in the divergence equation 
\begin{equation}
\label{chpt:divsa}
\partial_\mu A^\mu=\frac{3g^2_s}{16\pi^2}\epsilon_{\mu\nu\rho\sigma}
\mbox{Tr}_c\left({\bf F}^{\mu\nu}{\bf F}^{\rho\sigma}\right).
\end{equation}
   The factor of three originates from the number of flavors.
   In the large $N_c$ (number of colors) limit of
Ref.~\cite{tHooft:1973alw} the singlet axial-vector current is conserved, because the strong coupling constant
behaves as $g^2_s\sim N_c^{-1}$.

    In the quantized theory, the spatial integrals over the charge densities of 
the symmetry currents give rise to the charge operators $Q_{La}$, $Q_{Ra}$ ($a=1,\ldots, 8$), and $Q_V$.
     They are generators of the group $\text{SU}(3)_L\times\text{SU}(3)_R\times 
\text{U}(1)_V$, acting on the Hilbert space of QCD, and satisfy the commutation relations
\begin{subequations}
\begin{align}
\label{chpt:crqll}
[Q_{La},Q_{Lb}]&=if_{abc}Q_{Lc},\\
\label{chpt:crqrr}
{[Q_{Ra},Q_{Rb}]}&=if_{abc}Q_{Rc},\\
\label{chpt:crqlr}
{[Q_{La},Q_{Rb}]}&=0,\\
\label{chpt:crqlvrv} {[Q_{La},Q_V]}&=[Q_{Ra},Q_V]=0,
\end{align}
\end{subequations}
where the $f_{abc}$ are the totally antisymmetric structure constants of the Lie algebra of SU(3) \cite{Gell-Mann:1961omu}.
In the chiral limit, these operators are time independent, \ie, they commute with the Hamiltonian in the chiral limit,
\begin{equation}
\label{chpt:vrhq} [Q_{La},H^0_{\rm QCD}]=[Q_{Ra},H^0_{\rm QCD}]=[Q_V,H^0_{\rm QCD}]=0.
\end{equation}
It is convenient to consider the linear com\-bi\-na\-tions $Q_{Aa}\equiv Q_{Ra}-Q_{La}$ and $Q_{Va}\equiv Q_{Ra}+Q_{La}$, which transform as $Q_{Aa}\mapsto -Q_{Aa}$ and $Q_{Va}\mapsto Q_{Va}$ under parity.
The hadron spectrum can be organized in multi\-plets belonging to irreducible representations of SU(3)$_V$ with a given baryon number.
If not only the vector subgroup, but the full group were realized linearly by the spectrum of the hadrons, one would expect a so-called parity doubling of mass-degenerate states.
The absence of such a doubling in the low-energy spectrum is an indication that the $\text{SU}(3)_L\times\text{SU}(3)_R$ chiral symme\-try is dynamically broken in the ground state.
One then assumes that the axial generators $Q_{Aa}$ do not annihilate the ground state of QCD,
\begin{equation}
\label{chpt:QAavacuum}
Q_{Aa}|0\rangle\neq 0.
\end{equation}
As a consequence of the Goldstone theorem \cite{Goldstone:1962es}, each axial generator $Q_{Aa}$ not annihilating the ground state corresponds to a {\it massless} Goldstone-boson field $\phi_a$ with spin 0, whose symmetry properties are tightly connected to the generator in question.
The Goldstone bosons have the same transformation behavior under parity as the axial generators,
\begin{equation}
\label{chpt:parityphi}
\phi_a(t,\vec{x})\stackrel{P}{\mapsto}-\phi_a(t,-\vec{x}),
\end{equation}
\ie, they are pseudoscalars.
From Eqs.~\eqref{chpt:crqll} and \eqref{chpt:crqrr} one obtains $[Q_{Va},Q_{Ab}]=if_{abc}Q_{Ac}$ and thus the
Gold\-stone bosons  transform under the subgroup $\mbox{SU(3)}_V$, which leaves the vacuum invariant, as an octet:
\begin{equation}
\label{chpt:transformationphiqv}
[Q_{Va},\phi_b(x)]=if_{abc}\phi_c(x).
\end{equation}
The members of the pseudoscalar octet $(\pi,K,\eta)$ of the real world are identified as the Goldstone bosons of QCD and would be massless for massless quarks.

After turning on the quark masses in terms of the mass term
\begin{align*}
    {\cal L}_{\cal M}&=-\bar{q}{\cal M}q
    =-\left(\bar{q}_R {\cal M} q_L +\bar{q}_L {\cal M}^\dagger q_R\right),\\
    {\cal M}&=\text{diag}(m_u,m_d,m_s),
\end{align*}
the Goldstone bosons will no longer be massless (see below).
Moreover, the symmetry currents are no longer conserved.
In terms of the vector currents $V^\mu_a=R^\mu_a-L^\mu_a$ and the axial-vector currents $A^\mu_a=R^\mu_a-L^\mu_a$, the corresponding diver\-gences read  
\begin{align}
\label{chpt:dsva}
\partial_\mu V^\mu_a&=
i\bar{q}\left[{\cal M},\frac{\lambda_a}{2}\right]q,\quad
\partial_\mu A^\mu_a=
i\bar{q}\gamma_5\left\{\frac{\lambda_a}{2},{\cal M}\right\}q.
\end{align}

The properties of the currents corresponding to the approximate chiral symmetry of QCD can be summar\-ized as follows:
\begin{enumerate}
\item In the limit of massless quarks, the sixteen currents $L^\mu_a$
and $R^\mu_a$ or, alternatively, $V^\mu_a=R^\mu_a+L^\mu_a$ and $A^\mu_a=R^\mu_a-L^\mu_a$ are conserved.
   The same is true for the singlet vector current $V^\mu$, whereas the
singlet axial-vector current $A^\mu$ has an anomaly (see Eq.~\eqref{chpt:divsa}).
\item For any values of quark masses, the individual flavor currents $\bar{u}\gamma^\mu u$, $\bar{d}\gamma^\mu d$, and $\bar{s}\gamma^\mu s$ are always conserved in the strong interactions reflecting the flavor inde\-pen\-dence of the strong coupling and the diagonal
form of the quark-mass matrix.
   Of course, the singlet vector current $V^\mu$, being the sum of
the three flavor currents, is always conserved.
\item In addition to the anomaly, the singlet axial-vector current
has an explicit divergence due to the quark masses:
\begin{displaymath}
\partial_\mu A^\mu=2i\bar{q}\gamma_5{\cal M}q+
\frac{3 g^2_s}{16\pi^2}\epsilon_{\mu\nu\rho\sigma}\text{Tr}_c\left({\bf F}^{\mu\nu}{\bf F}^{\rho\sigma}\right).
\end{displaymath}
\item For equal quark masses, $m_u=m_d=m_s$, the eight vector currents
$V^\mu_a$ are conserved, because $[\lambda_a,{\mathbbm 1}]=0$.
   Such a scenario is the origin of the SU(3) symme\-try
originally proposed by Gell-Mann and Ne'eman \cite{Gell-Mann:1964ook}.
The eight axial-vector currents $A^\mu_a$ are not con\-served.
The divergences of the octet axial-vector currents of Eq.~\eqref{chpt:dsva} are proportional to pseudoscalar quadratic forms.
This can be interpreted as the micro\-scopic origin of the PCAC relation (partially con\-served axial-vector current) \cite{Gell-Mann:1964hhf,Adler:1968} 
which states that the diver\-gences of the axial-vector currents are pro\-portional to renormalized field operators representing the
lowest-lying pseudoscalar octet.
\item Taking $m_u=m_d\neq m_s$ reduces SU(3) flavor symme\-try to SU(2) isospin symmetry.
\item Taking $m_u\neq m_d$ leads to isospin-symmetry breaking.
\end{enumerate}
Besides the conservation properties of the currents, one may also calculate their commutators (current algebra), which may then be used to derive certain relations among QCD Green functions analogous to the Ward identities of Quantum Electro\-dynamics.
The set of all QCD Green functions involving color-neutral quark bilinears is very efficiently collected in a generating functional,
\begin{equation}
\label{chpt:gfgfvs} 
\exp\left(i Z_{\rm QCD}[ v,a,s,p]\right) 
=\langle 0|T\exp\left[i\int d^4\! x\, {\cal L}_{\rm ext}(x)\right]|0\rangle_0,
\end{equation}
where 
\cite{Gasser:1983yg,Gasser:1984gg}:
\begin{align}
\label{chpt:lqcds} {\cal L}_{\rm ext}&=\sum_{a=1}^8v^\mu_a\, \bar q\gamma_\mu\frac{\lambda_a}{2} q
+v^\mu_{(s)}\, \frac{1}{3}\bar q\gamma_\mu q + \sum_{a=1}^8a^\mu_a\,\bar q\gamma_\mu\gamma_5\frac{\lambda_a}{2} q
-\sum_{a=0}^8 s_a\, \bar q\lambda_a q +\sum_{a=0}^8 p_a\, i\bar q\gamma_5\lambda_a q
\nonumber\\
&=\bar{q}\gamma_\mu \left(v^\mu +\frac{1}{3}v^\mu_{(s)} +\gamma_5 a^\mu \right)q -\bar{q}(s-i\gamma_5 p)q,
\end{align}
where $\lambda_0=\sqrt{\frac{2}{3}}{\mathbbm 1}$.
    A particular Green function is then obtained through a partial functional 
derivative with respect to the corresponding external fields.
Note that both the quark field operators $q$ in ${\cal L}_{\rm ext}$ and the ground state $|0\rangle$ refer to the chiral limit, indicated by the subscript 0 in Eq.~\eqref{chpt:gfgfvs}.
   The quark fields are operators in the Heisenberg picture and
have to satisfy the equation of motion and the canonical anticommutation relations.
    From the generating func\-tional,  we can even obtain Green functions of the
``real world,'' where the  quark fields and the ground state are those with finite quark masses.
To that end one needs to evaluate the functional derivative of Eq.~\eqref{chpt:gfgfvs} at $s=\mbox{diag}(m_u,m_d,m_s)$.
    The chiral Ward identities result from an invariance of the generating functional 
of Eq.~\eqref{chpt:gfgfvs} under a {\it local} transformation of the quark fields and a simultaneous transformation of the external fields \cite{Gasser:1983yg,Gasser:1984gg},
\begin{subequations}
\begin{align}
\label{chpt:qltrans}
q_L&\mapsto\exp\left(-i\frac{\Theta(x)}{3}\right) V_L(x) q_L,\\
\label{chpt:qrtrans}
q_R&\mapsto\exp\left(-i\frac{\Theta(x)}{3}\right) V_R(x) q_R,
\end{align}
\end{subequations}
where $V_L(x)$ and $V_R(x)$ are independent space-time-dependent SU(3) matrices, provided the external fields are subject to the transformations
\begin{subequations}
\begin{align}
\label{chpt:lmutrans}
l_\mu&\mapsto V_L l_\mu V_L^{\dagger} +iV_L\partial_\mu V_L^{\dagger},\\
\label{chpt:rmutrans} 
r_\mu&\mapsto V_R r_\mu V_R^{\dagger}
+iV_R\partial_\mu V_R^{\dagger},\\
\label{chpt:vmutrans}
v_\mu^{(s)}&\mapsto v_\mu^{(s)}-\partial_\mu\Theta,\\
\label{chpt:spiptrans}
s+ip&\mapsto V_R(s+ip)V_L^{\dagger},\\
\label{chpt:smiptrans}
s-ip&\mapsto V_L(s-ip)V_R^{\dagger}.
\end{align}
\end{subequations}
The derivative terms in Eqs.~\eqref{chpt:lmutrans}-\eqref{chpt:vmutrans} serve the same purpose as in the construction of gauge theories, \ie, they cancel analogous terms originating from the kinetic part of the quark Lagrangian.

\section{Chiral perturbation theory for mesons}
\label{sec:MesonChPT}

Effective field theory (EFT) is a powerful tool for de\-scribing the strong 
interactions at low energies. 
    The essen\-tial idea behind EFT was formulated by Weinberg in Ref. \cite{Weinberg:1978kz} as follows:
    \begin{quote}
``...  if one writes down the most general possible Lagrangian,
including all terms consistent with assumed symmetry principles, and
then calculates matrix elements with this Lagrangian to any given
order of perturbation theory, the result will simply be the most
general possible S--matrix consistent with analyticity, perturbative
unitarity, cluster decomposition and the assumed symmetry
princi\-ples.''
\end{quote}
    In the present context, we want to describe the low-energy dynamics of QCD in terms
of its Goldstone bosons as effective degrees of freedom rather than in terms of quarks and gluons.
The resulting low-energy approximation is called (mesonic) chiral pertur\-bation theory (ChPT).
Its foundations are discussed in Ref.~\cite{Leutwyler:1993iq}.
    Since the interaction strength of the Goldstone bosons vanishes in the 
zero-energy limit and the quark masses are regarded as small perturbations around the chiral limit, the mesonic Lagrangian is organ\-ized in a simultaneous derivative and a quark-mass ex\-pan\-sion.
    This Lagrangian is expected to have exactly eight pseudo\-scalar degrees of freedom 
transforming as an octet under flavor $\mbox{SU(3)}_V$.
   Moreover, taking account of spontaneous symmetry breaking, the ground
state should only be invariant under $\mbox{SU(3)}_V\times\mbox{U(1)}_V$.
    Finally, in the chiral limit, we want the effective Lagrangian to be invariant 
under $\mbox{SU(3)}_L\times\mbox{SU(3)}_R\times\mbox{U(1)}_V$.

Our goal is to approximate the ``true'' generating functional $Z_{\rm QCD}[v,a,s,p]$ of Eq.~\eqref{chpt:gfgfvs} by a sequence $$Z^{(2)}_{\rm eff}[v,a,s,p] + Z^{(4)}_{\rm eff}[v,a,s,p] +\ldots\, ,$$ where the effective generating functionals are obtained using the effective field theory.
The rationale underlying this approach is the assumption that including all of the infinite number of effective functionals $Z^{(2n)}_{\rm eff}[v,a,s,p]$ will, at least in the low-energy region, generate a result which is equiv\-alent to that obtained from $Z_{\rm QCD}[v,a,s,p]$.
Because of spontaneous symmetry breaking, the chiral group $\mbox{SU(3)}_L\times\mbox{SU(3)}_R$ is realized nonlinearly on the Goldstone-boson fields \cite{Weinberg:1978kz,Coleman:1969sm}.
    We define the SU(3) matrix
\begin{equation}
\label{chpt:upar}
U(x)=\exp\left(i\frac{\phi(x)}{F_0}\right),
\end{equation}
where the field matrix $\phi$ is a Hermitian, traceless $3\times 3$ matrix,
 \begin{equation}
\label{chpt:phisu3}
\phi(x)=\sum_{a=1}^8 \phi_a\lambda_a  \equiv
\begin{pmatrix}
\pi^0+\frac{1}{\sqrt{3}}\eta &\sqrt{2}\pi^+&\sqrt{2}K^+\\
\sqrt{2}\pi^-&-\pi^0+\frac{1}{\sqrt{3}}\eta&\sqrt{2}K^0\\
\sqrt{2}K^- &\sqrt{2}\bar{K}^0&-\frac{2}{\sqrt{3}}\eta
\end{pmatrix},
\end{equation}
   and the parameter $F_0$ is the chiral limit of 
the pion-decay constant. 
Under {\it local} chiral transformations, $U(x)$ transforms as \cite{Gasser:1984gg}
\begin{equation}
    \label{chpt:Utrans}
    U(x)\mapsto V_R(x)U(x)V_L^\dagger(x).
\end{equation}
In particular, Eq.~\eqref{chpt:Utrans} implies for the field matrix $\phi$ the transformation behavior $\phi(x)\mapsto V\phi(x) V^\dagger$ under global flavor SU(3)$_V$, \ie, the Goldstone bosons indeed form an SU(3) octet [see Eq.~\eqref{chpt:transformationphiqv}].
The most general Lagrangian with the smallest (nonzero) number of exter\-nal fields is given by \cite{Gasser:1984gg}
\begin{equation}
\label{chpt:lol}
{\cal L}_2=\frac{F_0^2}{4}\mbox{Tr}[D_\mu U (D^\mu U)^\dagger]
+\frac{F^2_0}{4}\mbox{Tr}(\chi U^\dagger + U\chi^\dagger),
\end{equation}  
where 
\begin{subequations}
\begin{align}
\label{chpt:cdu}
D_\mu U&\equiv\partial_\mu U -i r_\mu U+iU l_\mu
\mapsto V_R D_\mu U V_L^\dagger,\\
\label{chpt:chi}
\chi&\equiv 2B_0(s+ip)\mapsto V_R\chi V_L^\dagger.
\end{align}
\end{subequations}
    If we denote a small four momentum as of ${\cal O}(q)$, the covariant derivative
counts as ${\cal O}(q)$ and $\chi$ as ${\cal O}(q^2)$ (see below), such that the lowest-order Lagrangian 
is of ${\cal O}(q^2)$, indicated by the subscript 2.
    Using the cyclic property of the trace, ${\cal L}_2$ is easily seen to be invariant 
under the transformations of Eqs.~\eqref{chpt:lmutrans}-\eqref{chpt:smiptrans} and  \eqref{chpt:Utrans}.
    Moreover, ${\cal L}_2$ is invariant under the simultaneous re\-place\-ments $U\leftrightarrow U^\dagger$, $l_\mu\leftrightarrow r_\mu$, and $\chi\leftrightarrow \chi^\dagger$.
    It is said to be of even intrinsic parity.

    At lowest order, the effective field theory contains two parameters $F_0$ and $B_0$.
    In order to pin down the meaning of $F_0$, we consider the axial-vector current $J^\mu_{Aa}$ associated with ${\cal L}_2$:
\begin{equation}
    \label{chpt:avcchpt}
    J^\mu_{Aa}=-i\frac{F^2_0}{4}
    \mbox{Tr}\left(\lambda_a\{U,\partial^\mu U^\dagger\}\right).
\end{equation}
Expanding $U$ in terms of the field matrix $\phi$, and using $\mbox{Tr}(\lambda_a\lambda_b)=2\delta_{ab}$ results in
\begin{equation}
J^\mu_{Aa}=-F_0\partial^\mu\phi_a+{\cal O}(\phi^3),
\end{equation}
from which we conclude that the axial-vector current has a nonvanishing matrix element when evaluated be\-tween the vacuum and a one-Goldstone-boson state:
\begin{equation}
\label{chpt:jamatrixelemtn}
\langle 0|J^\mu_{Aa}(x)|\phi_b(p)\rangle
=ip^\mu F_0\exp(-ip\cdot x)\delta_{ab}.
\end{equation}
Equation \eqref{chpt:jamatrixelemtn} holds at leading order (LO) in ChPT. It is the current-density analog of Eq.~\eqref{chpt:QAavacuum}, \ie, a nonvanishing value of $F_0$ is a necessary and suffi\-cient criterion for spontaneous symmetry breaking in QCD.
    
The expansion of the first term of Eq.~\eqref{chpt:lol} in the field matrix $\phi$ yields
\begin{equation}
    \label{chpt:l2kinexp}
    \frac{1}{4}\mbox{Tr}\left(\partial_\mu\phi\partial^\mu\phi\right)+\frac{1}{48 F^2}\mbox{Tr}\left(
[\phi,\partial_\mu \phi][\phi, \partial^\mu \phi]\right)+\ldots.
\end{equation}
The first term of Eq.~\eqref{chpt:l2kinexp} describes the kinetic term of the eight Goldstone bosons and the second term contributes to the scattering of Goldstone bosons.
The second term of Eq.~\eqref{chpt:lol} is an example how the explicit symmetry breaking by the quark masses is trans\-ferred from the QCD level to the EFT level.
    Both, ${\cal L}^0_{\rm QCD}+{\cal L}_{\rm ext}$ and ${\cal L}_2$ are invariant under 
{\it local} chiral transformations. 
    Inserting ${\cal L}_{\rm ext}={\cal L}_{\cal M}$ corresponds to
$s=\mbox{diag}(m_u,m_d,m_s)$ and it is the same $s$ that is to be used in the effective Lagrangian.
    The expansion of the $\chi$ term gives rise to
\begin{equation}
\label{chpt:lchiexp}
 F_0^2B_0(m_u+m_d+m_s)-\frac{B_0}{2}
 \mbox{Tr}\left(\phi^2{\cal M}\right)
 +2B_0\mbox{Tr}\left({\cal M}\phi^4\right)
 +\ldots
\end{equation}
Even though the first term of Eq.~\eqref{chpt:lchiexp} is of no dynamical significance for the interaction among the Goldstone bosons, it represents an interesting effect.
Its negative is the energy density of the vacuum, $\langle{\cal H}_{\rm eff}\rangle_{\rm min}$, which is shifted relative to the chiral limit because of the nonzero quark masses.
    We compare the partial derivative of $\langle{\cal H}_{\rm eff}\rangle_{\rm min}$ with
respect to (any of) the light-quark masses $m_l$ with the corresponding quantity in QCD,
\begin{equation}
    \left.\frac{\partial \langle 0|{\cal H}_{\rm QCD}|0\rangle}{\partial m_l}
    \right|_{m_u=m_d=m_s=0}=\frac{1}{3}\langle 0|\bar{q}{q}|0\rangle_0
=\frac{1}{3}\langle \bar{q}q\rangle_0,
\end{equation}
where $\langle\bar{q}{q}\rangle_0$ is the scalar singlet quark condensate.
With\-in the framework of the lowest-order effective Lagrangian, the constant $B_0$ is thus related to the scalar singlet quark condensate by
\begin{equation}
    \label{chpt:b0}
3 F^2_0B_0=-\langle\bar{q}q\rangle_0.
\end{equation}
For an overview of recent lattice QCD determinations of $\langle\bar{q}{q}\rangle_0$ see Ref.~\cite{FLAG}.
Because of the second term of Eq.~\eqref{chpt:lchiexp}, the Goldstone bosons are no longer massless.
If, for the sake of simplic\-ity, we consider the isospin-symmetric limit $m_u=m_d=\hat m$ (so that there is no $\pi^0$-$\eta$ mixing), we obtain for the masses of the Goldstone bosons, to lowest order in the quark masses (${\cal O}(q^2)$, denoted by the subscript 2),
\begin{subequations}
\begin{align}
    \label{chpt:mpi2}
    M^2_{\pi,2}&=2 B_0 \hat m,\\
    \label{chpt::mk2}
    M^2_{K,2}&=B_0(\hat m+m_s),\\
    \label{chpt:meta2}
    M^2_{\eta,2}&=\frac{2}{3} B_0\left(\hat m+2m_s\right).
\end{align}
\end{subequations}
These results, in combination with Eq.~\eqref{chpt:b0}, correspond to relations obtained in Ref.~\cite{Gell-Mann:1968hlm} and are referred to as the Gell-Mann, Oakes, and Renner relations.
Because of the on-shell condition $p^2=M^2$, Eqs.~\eqref{chpt:mpi2}-\eqref{chpt:meta2} justify the assignment $\chi={\cal O}(q^2)$.
Inserting the empirical values $M_\pi=135$ MeV, $M_K=496$ MeV, and $M_\eta=548$ MeV for the lowest-order predictions provides a first estimate for the ratio of the quark masses,
\begin{subequations}
\begin{eqnarray}
\frac{M^2_K}{M^2_\pi}=\frac{\hat m+m_s}{2\hat m}&\Rightarrow&\frac{m_s}{\hat m}=25.9,
\\
\frac{M^2_\eta}{M^2_\pi}=\frac{2m_s+\hat m}{3\hat m}&\Rightarrow&\frac{m_s}{\hat m}=24.3.
\end{eqnarray}
\end{subequations}

A remarkable feature of Eq.~\eqref{chpt:lol} is the fact that, once $F_0$ is known (from pion decay), chiral symmetry allows us to make absolute predictions about other pro\-cesses.
For example, the lowest-order results for the scattering of Goldstone bosons can be derived straight\-forwardly from the ${\cal O}(\phi^4)$ contribu\-tions of Eqs.~\eqref{chpt:l2kinexp} and \eqref{chpt:lchiexp}.
In particular, the $s$-wave $\pi\pi$-scattering lengths for the isospin channels $I=0$ and $I=2$ are obtained as \cite{Gasser:1983yg}
\begin{equation}
    \label{chpt:a00a02lo}
    a_0^0=\frac{7 M_\pi^2}{32 \pi F_\pi^2}=0.160,\quad
    a_0^2=-\frac{M_\pi^2}{16 \pi F_\pi^2}=-0.0456,
\end{equation}
   where we replaced $F_0$ by the physical pion-decay con\-stant and made use of the 
numerical values $F_\pi=92.2$ MeV and $M_\pi=M_{\pi^+}=139.57$ MeV. 
These results are identical with the current-algebra predictions of Ref.~\cite{Weinberg:1966kf}.
Actually, they serve as an illustration of the fact that the results of current algebra can (more easily) be repro\-duced from lowest-order perturbation theory in terms of an effective Lagrangian \cite{Weinberg:1966fm}---in the present case the lowest-order mesonic ChPT Lagrangian.

However, ChPT is much more powerful than the effective Lagrangians of the 1960s, which, by definition, were meant to be applied only in lowest-order per\-tur\-ba\-tion theory (see, e.g., the second footnote in Ref.~\cite{Schwinger:1967tc}).
In ChPT, a systematic improvement beyond the tree-level of the lowest-order Lagrangian of Eq.~\eqref{chpt:lol} is accom\-plished by calculating loop corrections in combination with tree-level contributions from Lagrangians of higher order.
For a long time it was believed that performing loop calculations using the Lagrangian of Eq.~\eqref{chpt:lol} would make no sense, because it is not renormalizable (in the traditional sense \cite{Dyson:1949ha}). 
However, as emphasized by Weinberg \cite{Weinberg:1978kz,Weinberg:1995mt}, the cancellation of ultraviolet diver\-gences does not really depend on renormalizability; as long as one includes every one of the infinite number of interactions allowed by symmetries, the so-called non-renormalizable theories are actually just as renormal\-izable as renormal\-izable theories
\cite{Weinberg:1995mt}. 
This still leaves open the question of how to organize a perturbative description of observables.
For that purpose, one needs a power-counting scheme to assess the importance of various diagrams calculated from the most general effec\-tive Lagrangian.
    Using Wein\-berg's power counting  
    scheme \cite{Weinberg:1978kz},
one may analyze the behavior of a given diagram of mesonic ChPT
under a linear re-scaling of all {\it external} momenta,
$p_i\mapsto t p_i$, and a quadratic re-scaling of the 
light-quark masses, $
m_l\mapsto t^2 m_l$, which, in terms of the Goldstone-boson masses, corresponds
to $M^2\mapsto t^2 M^2$.
   The chiral dimension $D$ of a given diagram with amplitude
${\cal M}(p_i,m_l)$ is defined by
\begin{equation}
\label{chpt:mr1} {\cal M}(tp_i, t^2 m_l)=t^D {\cal M}(p_i,m_l),
\end{equation}
where, in $n$ dimensions,
\begin{align}
D&=n N_L-2N_I+\sum_{k=1}^\infty 2k N_{2k}\label{chpt:mr2a}\\
&=2+(n-2) N_L+\sum_{k=1}^\infty 2(k-1) N_{2k}\label{chpt:mr2b}\\
&\geq\text{2 in 4 dimensions}.\nonumber
\end{align}
   Here, $N_L$ is the number of independent loops,
$N_I$ the number of internal Goldsone-boson lines, and $N_{2k}$ the number of
vertices originating from ${\cal L}_{2k}$.
   A diagram with chiral dimension $D$ is said to be of order ${\cal O}(q^D)$.
   Clearly, for small enough momenta and masses, dia\-grams with small $D$, such
as $D=2$ or $D=4$, should dominate.
   Of course, the re-scaling of Eq.~\eqref{chpt:mr1} must be viewed as
a mathematical tool.
   While external three-momenta can, to a certain extent, be made arbitrarily
small, the re-scaling of the quark masses is a theoretical instrument only.
   Note that, for $n=4$, loop diagrams are always suppressed due to the term
$2N_L$ in Eq.~\eqref{chpt:mr2b}.
   In other words, we have a perturbative scheme in terms of exter\-nal
momenta and masses which are small compared to some scale (here $4\pi F_0\approx 1$
GeV).

    The most general Lagrangian at ${\cal O}(q^4)$ was con\-structed by Gasser and Leutwyler \cite{Gasser:1984gg}
and contains twelve low-energy constants (LECs) ($L_1,\ldots,L_{10},H_1 H_2$),
\begin{align}
\label{chpt:l4gl}
    {\cal L}_4&=L_1\left\{\text{Tr}[D_{\mu}U (D^{\mu}U)^{\dagger}]\right\}^2+\ldots+H_2\text{Tr}\left(\chi\chi^{\dagger}\right).
\end{align}
   The numerical values of the low-energy constants $L_i$ are not determined by chiral symmetry.
   In analogy to $F_0$ and $B_0$ of ${\cal L}_2$ they are parameters containing information on the underlying dynamics.
For an extensive review of the status of these coupling constants, see Refs.~\cite{Bijnens:2014lea} as well as \cite{FLAG}.
 
\begin{figure}[t]
    \begin{center}
    \resizebox{0.4\textwidth}{!}{
    \includegraphics{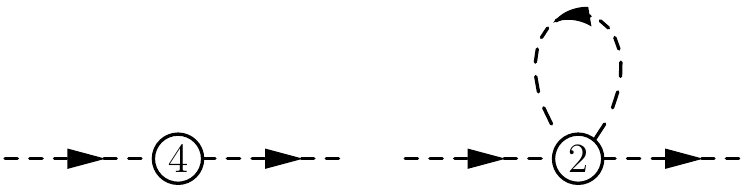}
    }
    \caption{\label{chpt:selfenergy} Self-energy diagrams at ${\cal O}(q^4)$.
   Vertices derived from ${\cal L}_{2n}$ are denoted by $2n$ in the
    interaction blobs.}
    \end{center}
\end{figure}

As an example of a one-loop calculation let us con\-sider the ${\cal O}(q^4)$ corrections to the masses of the Goldstone bosons. 
For that purpose one needs to evaluate the self-energy diagrams shown in Fig.~\ref{chpt:selfenergy}.
The corresponding expressions for the masses were first given in Ref.~\cite{Gasser:1984gg}, of which we show the squared pion mass as a representative example:
\begin{align}
\label{chpt:mpi24}
M^2_{\pi,4} &=
M^2_{\pi,2}\Bigg\{1+\frac{M^2_{\pi,2}}{32\pi^2F^2_0}
\ln\left(\frac{M^2_{\pi,2}}{\mu^2}\right)-\frac{M^2_{\eta,2}}{96\pi^2F^2_0}
\ln\left(\frac{M^2_{\eta,2}}{\mu^2}\right)\nonumber\\
&\quad+\frac{16}{F^2_0}\left[(2\hat m+m_s)B_0(2L^r_6-L^r_4)
+\hat mB_0(2L^r_8-L^r_5)\right]\Bigg\}.
\end{align}
    Because of the overall factor $M_{\pi,2}^2$, the pion stays 
mass\-less as $m_l\to 0$. 
This is, of course, what we expected from QCD in the chiral limit, but it is comforting to see that the self interaction in ${\cal L}_2$
(in the absence of quark masses) does not generate
Goldstone-boson masses at higher order.
    The ultraviolet divergences generated by the loop diagram of Fig.~\ref{chpt:selfenergy} are cancelled by a suitable adjustment of the
parameters of ${\cal L}_4$.
    This is Weinberg's argument on renormalizability at work; as long as one works
    with the most general Lagrangian all ultraviolet divergences can be absorbed in the parameters of the theory.
    At ${\cal O}(q^4)$, the squared Goldstone-boson masses contain terms
which are analytic in the quark masses, namely, of the form $m^2_l$
multiplied by the renormalized low-energy constants $L_i^r$.
   However, there are also non\-analytic terms  of the
type $m^2_l \ln(m_l)$---so-called  chiral logarithms---which do not involve new parameters.
   Such a behavior is an illustration of the mechanism found by Li and
Pagels \cite{Li:1971vr}, who noticed that a perturbation theory
around a symmetry, which is realized in the Nambu-Goldstone mode,
results in both analytic as well as nonanalytic expressions in
the perturbation.
Finally, by construction, the scale depen\-dence of the renormal\-ized coefficients $L_i^r$ entering Eq.~\eqref{chpt:mpi24} is  such that it cancels the scale dependence of the chiral logarithms \cite{Gasser:1984gg}.
   Thus, physical observables do not depend on the scale $\mu$.

    In terms of Fig.~\ref{chpt:selfenergy} and the result of Eq.~\eqref{chpt:mpi24},
we can also comment on the so-called chiral-symmetry-breaking scale
$\Lambda_\chi$ to be $\Lambda_\chi=4\pi F_0$ \cite{Manohar:1983md}.
   In a loop correction, every endpoint of an internal Goldstone-boson
line is multiplied by a factor of $1/F_0$, since the SU($3$) matrix of
Eq.~\eqref{chpt:upar} contains the Goldstone-boson fields in the
combination $\phi/F_0$.
   On the other hand, external momenta $q$ or Goldstone-boson masses
produce factors of $q^2$ or $M^2$ (see Eqs.~\eqref{chpt:l2kinexp} and
\eqref{chpt:lchiexp}).
   Together with a factor $1/(16\pi^2)$ remaining after integration   
in four dimensions they
combine to corrections of the order of $[q/(4\pi F_0)]^2$ for each independent loop.
Strictly speak\-ing, this particular integral generates an additional fac\-tor of 2, and the factor of $1/(16\pi^2)$ should be considered an estimate.

The Lagrangians discussed so far are of even intrinsic parity.
At ${\cal O}(q^4)$, they are incomplete, because they do not describe processes such as $K^+K^-\to\pi^+\pi^-\pi^0$ or $\pi^0\to\gamma\gamma$.
The missing piece is the effective Wess-Zumino-Witten (WZW) action \cite{Wess:1971yu,Witten:1983tw}, which accounts for the chiral anomaly.
The chiral anomaly results in the so-called anomalous Ward identities that give a particular form to the {\it variation} of the generating func\-tional \cite{Gasser:1983yg,Wess:1971yu}.
At leading order, ${\cal O}(q^4)$, and in the ab\-sence of external fields, the WZW action reads \cite{Wess:1971yu,Witten:1983tw},
\begin{align}
\label{WZWaction}
S^0_{\text{ano}}&=N_c\,S^0_{\text{WZW}},\nonumber\\
S^0_{\text{WZW}}&=-\frac{i}{240\pi^2}\int^1_0 d\alpha\int d^4x\epsilon^{ijklm}\text{Tr}(\mathcal{U}^L_i \mathcal{U}^L_j \mathcal{U}^L_k \mathcal{U}^L_l \mathcal{U}^L_m).
\end{align}
For the construction of the WZW action, the domain of definition of $U$ needs to be extended to a (hypothetical) fifth dimension,
\begin{equation}
\label{chpt:ualpha}
U(y)=\exp\left(i\alpha\frac{\phi(x)}{F_0}\right),
\end{equation}
where $y^i=(x^\mu,\alpha)$, $i=0,\ldots, 4$, and $0\leq\alpha\leq 1$.
    Minkowski space is defined as the surface of the five-dimensional space for $\alpha =1$.
The indices $i,\dots,m$ in Eq.~\eqref{WZWaction} run from 0 to 4, $y_4=y^4=\alpha$, $\epsilon_{ijklm}$ is the completely antisymmetric (five-dimensional) tensor with $\epsilon_{01234}=-\epsilon^{01234}=1$, and $\mathcal{U}^L_i=U^\dagger \partial U/\partial y^i$. 

In contrast to ${\cal L}_2$ and ${\cal L}_4$, $S^0_{\text{ano}}$ is of odd intrinsic parity, \ie, it changes sign under $\phi\to-\phi$.
Expanding the SU(3) matrix $U(y)$ in terms of the
Goldstone-boson fields, $U(y)={\mathbbm 1}+i\alpha \phi(x)/F_0+O(\phi^2)$, one obtains an infinite series of terms, each involving an odd number of
Goldstone bosons.
   For example, after some rearrange\-ments, the term with the smallest number of Goldstone bosons
reads
\begin{equation}
\label{chpt:swzw5phi}
S_{\rm WZW}^{5\phi}
=\frac{1}{240\pi^2 F^5_0}\int d^4\! x\,
\epsilon^{\mu\nu\rho\sigma}\mbox{Tr}(\phi\partial_\mu\phi\partial_\nu\phi
\partial_\rho\phi\partial_\sigma\phi).
\end{equation}
In particular, the WZW action without external fields involves at
least five Goldstone bosons \cite{Wess:1971yu}.
Again, once $F_0$ is known, after inserting $N_c=3$ one obtains a parameter-free prediction
for, e.g., the process $K^+K^-\to\pi^+\pi^-\pi^0$.
 
	In the presence of external fields, the anomalous action receives an additional term \cite{Witten:1983tw,Manes:1984gk,Bijnens:1993xi}
\begin{align}
S_{\text{ano}}=N_c(S^0_{\text{WZW}}+S^{\text{ext}}_{\text{WZW}})
\label{WZWextfields}
\end{align}
given by
\begin{align}
\label{chpt:swzwext}
S^{\text{ext}}_{\text{WZW}}&=-\frac{i}{48\pi^2}\int d^4x \epsilon^{\mu\nu\rho\sigma}\text{Tr}
\left[Z_{\mu\nu\rho\sigma}(U,l,r)
- Z_{\mu\nu\rho\sigma}(\mathbbm{1},l,r)\right].
\end{align}
where the explicit form of $Z_{\mu\nu\rho\sigma}(U,l,r)$ can be found in \cite{Manes:1984gk,Bijnens:1993xi}.
    At leading order, the action of Eq.~\eqref{chpt:swzwext} is responsible for the two-photon 
decays of the $\pi^0$ or the $\eta$.
    Quantum corrections to the WZW classical action do not renormalize the coefficient of
the WZW term.
    The counter terms needed to renormalize the one-loop singularities
at ${\cal O}(q^6)$ are of a conventional chirally in\-variant structure.
In the three-flavor sector, the most general odd-intrinsic-parity Lagrangian at ${\cal O}(q^6)$ con\-tains 23 independent terms \cite{Ebertshauser:2001nj,Bijnens:2001bb}.
For an overview of appli\-cations in the odd-intrinsic-parity sector, we refer to Ref.~\cite{Bijnens:1993xi}.

\section{ChPT for baryons}
\label{sec:BChPT}

ChPT was first extended to the baryon sector in Ref.~\cite{Gasser:1987rb}, which considered  a variety of matrix elements with single-nucleon incoming and outgoing states.
While the general approach is analogous to that in the mesonic sector, \ie, one considers the most general Lagrangian consistent with the symmetries of QCD and expands observables in a quark-mass and low-momentum expansion, the baryon sector exhibits some new features.
In particular, unlike the Goldstone-boson masses, the baryon masses do not vanish in the chiral limit. 
This has important consequences for obtaining a proper power counting of diagrams containing baryon lines and for the regular\-ization and renormalization of loop diagrams.
In the following we restrict the discussion to $\text{SU(2)}_L\times\text{SU(2)}_R$ chiral symmetry; for the extension to $\text{SU(3)}_L\times\text{SU(3)}_R$ see, e.g., the reviews of Refs.~\cite{Bernard:1995dp,Geng:2013xn} and references therein.
To construct the pion-nucleon Lagrangian, the proton ($p$) and neutron ($n$) fields are combined into an SU(2) doublet $\Psi$,
\begin{equation}
    \Psi = \begin{pmatrix} p \\ n \end{pmatrix}.
\end{equation}
The nucleon fields are chosen to transform under local $\text{SU(2)}_L\times\text{SU(2)}_R$ transformations as
\begin{equation}
    \Psi \to K(V_L,V_R,U)\Psi,
\end{equation}
where the SU(2) matrix $K$ depends on the left- and right-handed transformations as well as on the pion fields collected in $U$,
\begin{equation}
    K(V_L,V_R,U)  = \sqrt{V_RUV_L^\dagger}^{-1} V_R \sqrt{U}.
\end{equation}
The baryon Lagrangian also contains the covariant de\-riva\-tive of the nucleon field given by
\begin{equation}
    D_\mu \Psi = (\partial_\mu + \Gamma_\mu -i v^{(s)}_\mu)\Psi,
\end{equation}
with the connection \cite{Gasser:1987rb,Ecker:1994gg}
\begin{equation}
    \Gamma_\mu = \frac{1}{2} \left[ u^\dagger(\partial_\mu -i r_\mu)u + u (\partial_\mu - i l_\mu) u^\dagger \right],
\end{equation}
where $u^2 = U$, and the isoscalar vector field $v^{(s)}_\mu$.
Further, it is convenient to define 
\begin{equation}
    u_\mu = i \left[ u^\dagger(\partial_\mu -i r_\mu)u - u (\partial_\mu - i l_\mu) u^\dagger \right].
\end{equation}
The LO Lagrangian can be written as \cite{Gasser:1987rb}
\begin{equation}
\label{eq:LpiN1}
    \mathcal{L}_{\pi N}^{(1)} = \bar{\Psi} \left( i\slashed{D} -\mathtt{m} + \frac{\mathtt{g}_A}{2}\gamma^\mu \gamma_5 u_\mu \right) \Psi.
\end{equation}
It contains two LECs: $\mathtt{m}$ and $\mathtt{g}_A$. These correspond to the nucleon mass ($\mathtt{m}$) and the nucleon axial-vector coup\-ling constant ($\mathtt{g}_A$), both taken in the chiral limit. The corresponding physical values will be denoted as $m_N$ and $g_A$ in the following.
The superscript $(1)$ in Eq.~\eqref{eq:LpiN1} denotes that the Lagrangian is of first order in the power counting. 
While neither the nucleon energy nor the chiral-limit nucleon mass are small parameters, the combination $i\slashed{D} -\mathtt{m}$ can be assumed to be a small quan\-tity as long as the nucleon three-momentum is ${\cal O}(q)$.

This Lagrangian can be used to calculate the first loop contribution to the nucleon mass. The power count\-ing predicts this contribution to be ${\cal O}(q^3)$. 
However, the application of dimensional regularization and the minimal subtraction scheme of ChPT ($\widetilde{\text{MS}}$) as used in the meson sector results in terms that are of lower order than predicted by the power counting. 
Analogous issues also arise for other observables and higher-order contributions.
The authors of Ref.~\cite{Gasser:1987rb} pointed out that the failure of the power counting is related to the regular\-ization and renormalization schemes and that the ``same phenomenon would occur in the meson sector, if one did not make use of dimensional regularization.''
Several methods to address the power counting issue have been proposed \cite{Jenkins:1990jv,Ellis:1997kc,Becher:1999he,Gegelia:1999gf,Gegelia:1999qt,Fuchs:2003qc}.

One commonly used method is Heavy Baryon ChPT (HBChPT) \cite{Jenkins:1990jv}, which was inspired by Heavy Quark Effective Theory \cite{Georgi:1990um,Eichten:1989zv}.
Because the nucleon mass is large compared to the pion mass, an additional expan\-sion of the pion-nucleon Lagrangian is performed in inverse powers of the nucleon mass. 
In this formalism, application of dimensional regularization in combination with $\widetilde{\text{MS}}$ to loop diagrams, as in the meson sector, leads to a consistent power counting, connecting the chiral to the loop expansion.
The heavy-baryon Lagrangian up to and including order $q^4$ is given in Ref.~\cite{Fettes:2000gb}.
For an introduction to, and applications of, this method see, e.g., Refs.~\cite{Bernard:1995dp,Scherer:2002tk}. 

While the heavy-baryon formalism makes it possible to use techniques from the meson sector, the additional expansion in powers of the inverse nucleon mass results in a large number of terms in the higher-order Lagrangians.
Some of the higher-order terms are related to those at lower orders by Lorentz invariance \cite{Luke:1992cs}. 
Calculated amplitudes can be expressed in Lorentz-invariant forms, but Lorentz invariance is not manifest throughout intermediate steps of the calculations.
Fur\-ther, issues with analyticity arise in some specific cases because the heavy-baryon expansion results in a shift of the poles in the nucleon propagator \cite{Becher:1999he}.

A manifestly Lorentz-invariant approach to baryon ChPT that addresses these issues was formulated in Ref.~\cite{Becher:1999he}, referred to as infrared regularization.
While infrared regularization also uses dimensional regular\-ization, the renormalization procedure is different from minimal subtraction. 
Loop integrals are separated into infrared-singular and infrared-regular parts. 
The infra\-red-singular parts contain the same infrared singularities as the original integral and they satisfy the power count\-ing.
The infrared-regular parts are analytic in small parameters for arbitrary spacetime dimensions and con\-tain the power-counting-violating terms. 
Since the infra\-red-regular parts are analytic, they can be absorbed in the LECs of the baryon Lagrangian.
Infrared regular\-ization in its original formulation was applicable to one-loop diagrams. 
It has been widely used in the calculation of baryon properties, see, e.g., Ref.~\cite{Bernard:2007zu} for a review.

The expansion of the infrared-regular parts in small parameters contains not only the terms violating the power counting, but also an infinite set of terms that satisfy the power counting. 
The extended on-mass-shell (EOMS) scheme \cite{Fuchs:2003qc} provides a method to isolate the terms that violate the power counting and to absorb only these terms in the LECs of the Lagrangian. 
The EOMS scheme was also shown to be applicable to multi-loop diagrams \cite{Schindler:2003je} and diagrams containing particles other than pions and nucleons \cite{Fuchs:2003sh}.
By reformulating infrared regularization analogously to the EOMS scheme \cite{Schindler:2003xv}, it can be applied beyond one-loop pion-nucleon diagrams \cite{Schindler:2003je}; see also Ref.~\cite{Bruns:2004tj} for a different extension of infrared regularization.

The nucleon mass presents an example of the appli\-cation of baryon ChPT. It has been determined to one-loop order in several approaches, 
including HBChPT \cite{Steininger:1998ya}, infrared regularization \cite{Becher:1999he}, and the EOMS scheme \cite{Fuchs:2003qc}.
Up to and including order $q^3$, the chiral expansion of the nucleon mass is given by 
\begin{equation}
    \label{chpt:mN}
    m_N = \mathtt{m} -4 c_1 M^2 - \frac{3 \mathtt{g}_A^2}{32 \pi F^2}M^3 +\ldots,
\end{equation}
where $F$ denotes the pion-decay constant in the two-flavor chiral limit, $F_\pi=F[1+{\cal O}(\hat m)]=92.2$ MeV and $M^2=2B\hat m$ is the lowest-order expression for the squared pion mass. 

The result of Eq.~\eqref{chpt:mN} exhibits some general features of baryon ChPT: The expansion contains not just even powers in the small parameter $q$ like the meson sector, but also odd powers. As a result, the convergence of chiral expansions is expected to be slower in the baryon sector. 
The second-order contribution is proportional to the LEC $c_1$ from the second-order Lagrangian. On the other hand, the coefficient of the nonanalytic term proportional to $M^3$ is given entirely in terms of the LO LEC $\mathtt{g}_A$ and $F$. Similar features also appear at higher orders.
The general form of the chiral expansion of the nucleon mass to higher orders is given by
\begin{equation}
\label{eq:mN6}
\begin{split}
    m_N & = \mathtt{m} + k_1 M^2 + k_2 M^3 + k_3 M^4 \ln\left(\frac{M}{\mu}\right) + k_4 M^4 
    + k_5 M^5 \ln\left(\frac{M}{\mu}\right) + k_6 M^5 \\ 
    &
    + k_7 M^6 \ln^2\left(\frac{M}{\mu}\right) + k_8 M^6 \ln\left(\frac{M}{\mu}\right)+ k_9 M^6 + \ldots,
\end{split}
\end{equation}
where $\mu$ is the renormalization scale and the ellipsis denotes higher-order terms.
The coefficients $k_i$ are linear combinations of various LECs. $k_1$ through $k_4$ can be determined by considering at most one-loop diagrams, while $k_5$ through $k_9$ receive contributions from two-loop diagrams.
Using estimates of the LECs entering the $k_i$, Ref.~\cite{Fuchs:2003kq} estimated the nucleon mass in the chiral limit from an EOMS calculation to order $q^4$ to be
\begin{equation}
\begin{split}
    \mathtt{m} & = [938.3-74.8+15.3+4.7-0.7]\, \text{MeV} \\
    & = 882.8 \,\text{MeV}.
\end{split}
\end{equation}
Two-loop contributions to order $q^5$ were considered in Ref.~\cite{McGovern:1998tm}, while Refs.~\cite{Schindler:2006ha,Schindler:2007dr} determined $m_N$ to order $q^6$.
Because several currently undetermined LECs enter the expressions for several of the higher-order $k_i$, no reliable estimate of the complete two-loop contributions is possible.
However, the coefficient $k_5$ of the leading nonanalytic contribution at order $q^5$ only depends on $\mathtt{g}_A$ and the pion-decay constant $F$ and can therefore be compared to lower-order terms.
At the physical pion mass and with $\mu = m_N$, $k_5 M^5 \ln(M/m_N) = -4.8\,\text{MeV}$.

Chiral expansions like that of Eq.~\eqref{eq:mN6} are also important at nonphysical pion masses in the extrapolation of lat\-tice QCD results (for an introduction see, e.g., Ref.~\cite{Golterman:2009kw}).
The fifth-order term $k_5 M^5 \ln(M/m_N)$ becomes as large as the third-order term $k_2 M^3$, where $k_2$ also only de\-pends on $\mathtt{g}_A$ and $F$, for a pion mass of about 360 MeV.
While this comparison includes only one part of the two-loop contributions, it indicates a limit to the applicability of the power counting.
This estimate agrees with others found using different methods in Refs.~\cite{Meissner:2005ba,Djukanovic:2006xc}.

Even though the nucleon mass is a static quantity, it is not entirely surprising that a combined chiral and momentum expansion in the baryon sector does not converge well for energies beyond about $300\,\text{MeV}$. 
This roughly corresponds to the mass gap between the nu\-cleon and the $\Delta(1232)$ resonance.
At the physical point, treating the $\Delta$ as an explicit degree of freedom has limited impact on the nucleon mass \cite{Bernard:2003xf,Hacker:2005fh}.
However, the $\Delta(1232)$ also couples strongly to the $\pi N$ channel and has relatively large photon decay amplitudes, resulting in important contributions to processes such  as pion-nucleon scattering, Compton scattering, and electromagnetic pion production.
These issues were already pointed out in Ref.~\cite{Jenkins:1990jv}, which advocated for treating $\Delta$ degrees of freedom as dynamic.
In baryon ChPT with only pions and nucleons as degrees of free\-dom, effects of the $\Delta(1232)$ enter implicitly through the values of the LECs.
However, these contributions can be proportional to powers of $M/\delta$, where $\delta = (m_\Delta - m)$.
This ratio is small as the quark masses approach the chiral limit, but it is a rather large expansion parameter at the physical values, especially when combined with the strong coupling of the $\Delta$.
By formulating a theory that also includes the $\Delta$ as an active degree of freedom, one hopes to improve the convergence of the pertur\-bative expansion and potentially to increase the kin\-ematic range of applicability.

The inclusion of $\Delta$ degrees of freedom poses addi\-tional challenges to the construction of the most general Lagrangian and to the power counting.
The covariant description of spin-$\frac{3}{2}$, isospin-$\frac{3}{2}$ fields introduces unphysi\-cal degrees of freedom \cite{Rarita:1941mf,Moldauer:1956zz}. 
For the free Lagrangian, these can be eliminated by subsidiary equations and projection operators.
The correct number of degrees of freedom also has to be preserved when including interactions with pions, nucleons, and external fields. 
Various approaches addressing this issue have been con\-sidered, see, e.g., Refs.~\cite{Nath:1971wp,Tang:1996sq,Hemmert:1997ye,Pascalutsa:1998pw,Wies:2006rv,Krebs:2009bf}.

The main issue for the power counting is how to count the $\Delta$-nucleon mass difference $\delta$.
In one version of the power counting \cite{Hemmert:1997ye}, it is a small quantity of the same order as the pion mass, $\delta \sim {\cal O}(q)$. 
In a different approach \cite{Pascalutsa:2002pi}, it is argued that (for physical quark masses) $M_\pi < \delta$ and that $M_\pi/\delta \sim \delta/\Lambda$, where $\Lambda\sim 1\,\text{GeV}$ is the breakdown scale of the EFT. Denoting $\bar{\delta} \equiv \delta/\Lambda$ implies that $M_\pi/\Lambda \sim \bar{\delta}^2$, \ie, the pion mass is of higher order than the $\Delta$-nucleon mass difference in this power count\-ing. 

\section{Conclusions}

Over the last few decades, ChPT has developed into a mature and comprehensive approach to the  low-energy interactions between Goldstone bosons, nucleons, and external fields, with numerous successful applications.
ChPT has played an important role in interpreting lat\-tice QCD calculations performed at unphysical pion masses. 
It has also served as a prototype for semi-phenom\-enological approaches in other systems.
The application of ChPT methods to the interactions be\-tween two and more nucleons is discussed in the contri\-bution by Epelbaum and Pastore.

\begin{acknowledgments}
This work was supported by the U.S.~Department of Energy, Office of Science, Office of Nuclear Physics, under Award Number DE-SC0019647 (MRS).
\end{acknowledgments}

%

\end{document}